\documentclass{article}
\usepackage{authblk}
\usepackage{graphicx}
\usepackage{amsmath,amssymb,amsthm}
\usepackage{slashed}
\usepackage{appendix}
\usepackage[left=0.8in,right=0.8in,top=1in,bottom=1in]{geometry}
\begin{document}
	\title{Semileptonic Decay of $\Omega_c^0 \to \Xi^- l^+ \nu_l$ From Light-Cone Sum Rules}
	\author{Hui-Hui Duan\footnote{duanhuihui19@nudt.edu.cn}}
	\author{Yong-Lu Liu\footnote{yongluliu@nudt.edu.cn}}
	\author{Ming-Qiu Huang\footnote{Corresponding author: mqhuang@nudt.edu.cn}}
	\affil{Department of Physics, College of Liberal Arts and Sciences, National University of Defense Technology, Changsha 410073, Hunan, People's Republic of China}
	\renewcommand*{\Affilfont}{\small\it}
	\date{\today}
    \maketitle
	\begin{abstract}
	The weak decay process of $\Omega_c$ to $\Xi$ is calculated in the method of QCD light-cone sum rule. The decay width of $\Omega_c^0 \to \Xi^- l^+ \nu_l$ and its decay branching ratio are also calculated with the form factors from this work's calculation. To the twist-6 distribution amplitudes, the form factors $f_1=0.66\pm0.02, f_2=-0.76\pm0.03, g_1=0.06\pm0.01$ and $g_2=-0.44\pm0.01$ are given at zero recoil point. The result of the semileptonic decay width of $\Omega_c^0 \to \Xi^-l^+\nu_l$ is $\Gamma=(7.51\pm0.36)\times10^{-15}~{\rm{GeV}}$ , and the prediction of the decay branching ratio $Br(\Omega_c^0\to\Xi^-l^+\nu_l)=(3.06\pm0.15)\times10^{-3}$. These results fit well with other works, and the decay width and branching ratio are improved. This not too small branching ratio gives a good direction to explore this decay channel in the future experiments.
	\end{abstract}
%	 \pacs{13.30.Ce, 11.55.Hx, 12.38.Lg, 14.20.Lq}
	\section{Introductions} \label{sec:I} 
   In recent years, many new experimental results of $\Omega_c^0$ baryon have been developed. The lifetime has been updated by LHCb and a new value $\tau(\Omega_c^0 )=(268\pm24\pm10\pm2)\times10^{-15}s$~\cite{Aaij:2018} is given, about five times larger than the old measurements~\cite{Link:2003,Adamivich:1995,Frabetti:1995}. And also, five new narrow $\Omega_c^0$ states are reported by LHCb in 2017~\cite{Aaij:2017}, and confirmed by $e^+ e^-$ collisions on Belle~\cite{Yelton:2018}. These new discoveries enrich the nature of $\Omega_c^0$ baryon greatly. Two of the important properties of $\Omega_c^0$ baryon are its decay properties, strong and weak decays, but there is no strong decay observed in experiment until now. The weak decay channels are the main decay channels of $\Omega_c^0$. The PDG listed fourteen Cabibbo-favored weak decay channels up to now~\cite{Zyla:2020}, among which there are thirteen non-leptonic and only one semileptonic decay channel is observed. For investigation of the transition from $\Omega_c^0$ baryon to other lighter baryons, the simplicity object is the semileptonic weak decay.
   
   One of the established semileptonic decay mode is the channel $\Omega_c^0\to\Omega^- e^+ \nu_e$~\cite{Ammar:2002}.In this channel, the decay mode is charm quark decaying to strange quark and radiative positron and neutrino. For charm quark, the other possible decay channel $c\to dl^+\nu_l$ is not forbidden in the standard model, which was observed in $D^+\to \eta \mu^+ \nu_\mu$ semileptonic decay by BESIII collaboration recently and also gives a new value of CKM element $|V_{cd}|=0.242$~\cite{Ablikim:2020}. Therefore, the study of semileptonic decay of charm baryon provides an additional source of information for determining the CKM matrix elements of charm quark as well as exploring the internal dynamics of systems containing heavy-light quarks. 
  
   The transition from $\Omega_c^0$ baryon to $\Xi$ baryon has been studied with many theoretical approaches, such as nonrelativistic quark model~\cite{Rerez:1989}, heavy quark effective theory~\cite{Chen:1997,Dhir:2015}, light-front quark model~\cite{Zhao:2018}, and MIT bag model~\cite{Hu:2020}. With the development of experimental results of the mass, lifetime, and other parameters, the new calculation and prediction are necessary. In this work, we will study the transition with the method of light-cone sum rules~\cite{Chernyak:1984,Balitsky:1989,Braun:1989,Chernyak:1990,Huang:2004,Liu:2009lc}, which is based on QCD sum rules~\cite{Shifman:1979,Zhang:2008,Wang:2010,Agaev:2017o}. We will evaluate our work from baryon and weak currents and estimate the semileptonic decay width with the new result of $\Omega_c^0$ baryon lifetime. This method has been used to study the strong decay properties of excited $\Omega_c^0$ baryons~\cite{Chen:2017,Agaev:2017i}.
   
   After the introduction in Sec.~\ref{sec:I}, the details of the light-cone sum rules derivation of the semileptonic decay of $\Omega_c^0\to\Xi^-l^+\nu_l$ are given, and also the sum rules of the form factors are given in Sec.~\ref{sec:II}. Sec.~\ref{sec:III} is the numerical analysis of the four form factors of transition matrix element of $\Omega_c^0\to\Xi$. Conclusions are given in Sec.~\ref{sec:IV}. In the Appendix, the explicit expressions of $\Xi$ baryon distribution amplitudes are listed.

    \section{Light-cone sum rules of semileptonic decay $\Omega_c^0 \to \Xi^- l^+ \nu_l$.} \label{sec:II}

	Weak decay dynamics can be investigated by the weak decay effective Hamiltonian~\cite{Buchalla:1996}, and the semileptonic decay can be processed as that of the rare decay of B mesons~\cite{Hurth:2003}. For the $c\to dl^+\nu_l$ decay mode, the Hamiltonian can be wrote with the form~\cite{Li:2017}:
	\begin{gather}
	   \mathcal{H}_{eff}=\frac{G_F}{\sqrt{2}}|V_{cd}|O_1O_2,	
	\end{gather}
    where $G_F$ is fermi constant, $|V_{ud}|$ and $|V_{cd}|$ are the CKM matrix elements, $O_1=\bar{d}\gamma_\mu(1-\gamma_5)c,$ $O_2=\bar{l}\gamma^\mu(1-\gamma_5)\nu_l$ is quark and lepton current respectively.
   
     The light-cone sum rule starts from the current algebra structure of hadrons, and is evaluated with analytical methods, and then gives the hadron transition matrix element with form factors in theoretical expressions. This method uses two kinds of representations, hadronic and QCD theoretical representations. They are connected through quark hadron duality by using dispersion relations.

	The calculation of decay properties of heavy baryons need to know the decay matrix element. Matrix element of heavy baryon decay to light baryon $\Omega_c \to \Xi$ can be parameterized as six form factors as follows, 
  	   \begin{align}
    	\langle \Omega_c(P')&|j_\nu|\Xi(p)\rangle=\bar{u}_{\Omega_c}(P')[f_1\gamma_\nu- i \frac{f_2}{M_{\Omega_c}} \sigma_{\nu\mu}q^\mu - \notag \\& \frac{f_3}{M_{\Omega_c}}q_\nu-(g_1\gamma_\nu+i \frac{g_2}{M_{\Omega_c}} \sigma_{\nu\mu} q^\mu -\frac{g_3}{M_{\Omega_c}}q_\nu)\gamma_5]u_{\Xi}(p), \label{transition element}
	   \end{align}	
    where $f_i(g_i),i=1,2,3$ are the weak decay form factors, $M_{\Omega_c}$ the mass of $\Omega_c$ baryon, $P'=p-q$, q is the momentum transfer, $u_{\Omega_c}$ and $u_\Xi$ are the spinors of $\Omega_c$ and $\Xi$, respectively.
	
	In order to obtain the light-cone sum rules of these form factors, one begins with the two-point correlation function sandwiched between vacuum and final baryon state
       \begin{gather}
    	T_\nu (p,q)=i\int d^4x e^{iq \cdot x}\langle 0|T\{j_{\Omega_c}(0)j_\nu (x)\}|\Xi(p)\rangle  \label{correlation function}
       \end{gather}
	 where the $j_{\Omega_c}(0)$ and $j_\nu (x)$ are heavy baryon $\Omega_c$ current and weak decay current respectively. In this study, the Ioffe-type current is chosen for this computation
	   \begin{gather}
    	j_{\Omega_c}(x)=\epsilon_{ijk}(s^{iT}(x)C\gamma_\mu s^j(x))\gamma^\mu\gamma_5c^k(x)   \label{baryon current}
	   \end{gather} 
	 and the weak decay current
	   \begin{gather}
	    j_\nu(x)=\bar{c}(x)\gamma_\nu(1-\gamma_5)d(x).    \label{weak current}
	   \end{gather}
	
	 By defining a light-cone vector $z^\nu$ satisfying the condition $z^2=0$ to simplify the next calculations, which gives 
	    \begin{gather}
	     z^\nu T_\nu (p,q)=iz^\nu \int d^4x e^{iq \cdot x}\langle 0|T\{j_{\Omega_c}(0)j_\nu (x)\}|\Xi(p)\rangle.  \label{correlation function 2}
	    \end{gather}

	  The coupling constant of the baryon $\Omega_c$ is defined by the following transition matrix element formula
	    \begin{gather}
	\langle 0|j_{\Omega_c} |\Omega_c \rangle =  f_{\Omega_c} u_{\Omega_c} (P').
	    \end{gather}
	
	 With using of the completeness relation of $\Omega_c$ baryon, the hadronic representation can be expressed as
    
%    \begin{widetext}
    	\begin{align}
      z^\nu T_\nu(p,q)=&\frac{f_{\Omega_c}}{M_{\Omega_c}^2-P'^2}\{2f_1p\cdot z-M_\Xi f_1\slashed{z}-\frac{M_\Xi}{M_{\Omega_c}}f_2z\cdot q+\frac{f_2}{M_{\Omega_c}}2p\cdot z\slashed{q}-\frac{f_2}{M_{\Omega_c}}2 p\cdot q\slashed{z}+\frac{M_\Xi}{M_{\Omega_c}}f_2\slashed{z}\slashed{q}-\frac{M_\Xi}{M_{\Omega_c}}f_3q\cdot z\nonumber \\& -2g_1p\cdot z\gamma_5-M_\Xi g_1\slashed{z}\gamma_5+\frac{M_\Xi}{M_{\Omega_c}}g_2z\cdot q\gamma_5+\frac{g_2}{M_{\Omega_c}}2p\cdot z\slashed{q}\gamma_5-\frac{g_2}{M_{\Omega_c}}2p\cdot q\slashed{z}\gamma_5-\frac{M_\Xi}{M_{\Omega_c}}g_2\slashed{z}\slashed{q}\gamma_5+\frac{M_\Xi}{M_{\Omega_c}}g_3z\cdot q\gamma_5\nonumber \\&-2f_1q\cdot z+f_1\slashed{z}\slashed{q}+\frac{f_2}{M_{\Omega_c}}z\cdot q\slashed{q}-\frac{2f_2}{M_{\Omega_c}}q\cdot z\slashed{q}+\frac{f_2}{M_{\Omega_c}}q^2\slashed{z}+\frac{f_3}{M_{\Omega_c}}q\cdot z\slashed{q}\nonumber\\& +2g_1q\cdot z\gamma_5-g_1\slashed{z}\slashed{q}\gamma_5+\frac{g_2}{M_{\Omega_c}}z\cdot q\slashed{q}\gamma_5-\frac{g_2}{M_{\Omega_c}}2q\cdot z\slashed{q}\gamma_5+\frac{g_2}{M_{\Omega_c}}q^2\slashed{z}\gamma_5-\frac{g_3}{M_{\Omega_c}}z\cdot q\slashed{q}\gamma_5\nonumber\\&+M_{\Omega_c}f_1\slashed{z}-f_2z\cdot q+f_2\slashed{z}\slashed{q}-f_3q\cdot z\nonumber\\&-M_{\Omega_c}g_1\slashed{z}\gamma_5-g_2z\cdot q\gamma_5+g_2\slashed{z}\slashed{q}\gamma_5-g_3z\cdot q\gamma_5\}\Xi(p)+\cdots. 
        \end{align}
%       \end{widetext} 
        
	 In the above equation, the relationship $\sum_s u_{\Omega_c} (P') \bar{u}_{\Omega_c}(P') =(\slashed{P'}+M_{\Omega_c})$ has been used. $M_\Xi$ is the mass of $\Xi$ baryon and $M_{\Omega_c}$ is the mass of $\Omega_c$ baryon. The ellipsis stands for the continuum higher than threshold $s_0$. We set $q\cdot z=0$ in the following calculation because the small value on the light cone of the transfer momentum. Only the light leptons $e^\pm$ and $\mu^\pm$ are considered in this work. Their masses are very small in this system. In the zero masses limit, with the relation $q_\mu\bar{l}\gamma^\mu(1-\gamma_5)=0$, the $f_3$ and $g_3$ terms do not contribute. One can omit them and the corresponding terms in our following calculations~\cite{Huang:2004}.
    
     The next work is to derive the QCD theoretical side of the correlation function. In this work, the light-cone sum rule is used to express the QCD theoretical formalism. For this purpose, we expand the correlation function on the light-cone with the LCDAs of $\Xi$ baryon~\cite{Liu:2009lc}. Substituted the heavy baryon current (\ref{baryon current}) and the weak decay current (\ref{weak current}) into the correlation function (\ref{correlation function 2}), after contracting the heavy quark, we get the correlation function
        \begin{gather}
        z^\nu T_\nu=\int d^4 x e^{iq\cdot x} \int \frac{d^4 x}{(2\pi)^4} \frac{e^{ik\cdot x}}{k^2-m_c^2} (C \gamma_{\sigma})_{\alpha \beta}[\gamma_\sigma \gamma_5 (\slashed{k}+m_c) \slashed{z} (1-\gamma_5)_{\rho\gamma}]\langle 0 |\epsilon_{ijk} s^{iT}_\alpha (0) s^j_\beta (0) d^k_\gamma (x) |\Xi (p)\rangle.
        \end{gather}
        
   In these calculations, the LCDAs transformation relations are used, and the LCDAs of $\Xi$ has been given in~\cite{Liu:2009lc}. The LCDAs and other formulas needed in the calculations are also given in Appendix A, where only the parameters needed are displayed, and the completeness form can be found in~\cite{Liu:2009lc, Liu:2009da}. After the standard procedure of light-cone sum rule calculations and performing Borel transformations both on the two sides of hadron and theoretical representation, by choosing the structures $1, \slashed{z}\slashed{q}, \gamma_5$ and $\slashed{z}\slashed{q}\gamma_5$ in our sum rules. The final light-cone sum rules of these form factors $f_i (g_i) (i=1,2)$  are as follows:
  
%  \begin{widetext}
  	 Form factor $f_1(q^2)$
   \begin{align}	
  f_{\Omega_c}f_1 e^{-M_{\Omega_c}^2/M_B^2}=&\int_{\alpha_{30}}^{1}\{\frac{1}{\alpha_3}\rho^1_1(\alpha_3)-\frac{1}{\alpha_3^2 M_B^2}\rho^1_2(\alpha_3,q^2)  -\frac{1}{\alpha_3^3M_B^4}\rho^1_3(\alpha_3)\} e^{-s/M_B^2} \nonumber\\& -\frac{1}{\alpha_{30}^2 M_{\Xi}^2-q^2+m_c^2}[\rho^1_2(\alpha_{30},q^2)+\frac{1}{\alpha_{30}M_B^2}\rho^1_3(\alpha_{30})] e^{-s_0/M_B^2} \nonumber\\& +\frac{\alpha_{30}^2}{\alpha_{30}M_{\Xi}^2-q^2+m_c^2}[\frac{d}{d\alpha_{30}}\frac{\rho^1_3(\alpha_{30})}{\alpha_{30}(\alpha_{30}^2M_{\Xi}^2-q^2+m_c^2)}] e^{-s_0/M_B^2},
   \end{align}
   
   where the $\rho^1_i(i=1,2,3)$ are
   \begin{align}
   	\rho^1_1(\alpha_3)&=B_1(\alpha_3)m_c+\alpha_3 B_3(\alpha_3)M_\Xi,\\
   	\rho^1_2(\alpha_3,q^2)&=B_2(\alpha_3)[\alpha_3(\alpha_3-1)M_\Xi^3-M_\Xi m_c^2-\alpha_3M_\Xi q^2+\alpha_3M_{\Omega_c}^2M_\Xi-\alpha_3M_\Xi^2m_c]\nonumber \\&+\frac{1}{2}\alpha_3M_\Xi^2m_cB_4(\alpha_3)+\frac{1}{2}(\alpha_3M_\Xi^2m_c+\alpha_3^2M_\Xi^3)B_5(\alpha_3),\\
   	\rho^1_3(\alpha_3)&=\alpha_3^2M_\Xi^4m_c B_6(\alpha_3).
   	\end{align}
   
      Form factor $f_2(q^2)$
   \begin{align}
  f_{\Omega_c}(f_1&+\frac{M_{\Omega_c}+M_{\Xi}}{M_{\Omega_c}}f_2)  e^{-M_{\Omega_c}^2/M_B^2}=\int_{\alpha_{30}}^{1}d\alpha_3\{-\frac{1}{\alpha_3}\rho^2_1(\alpha_3) \nonumber\\& -\frac{1}{2\alpha_3^2M_B^2}\rho^2_2(\alpha_3)+\frac{1}{\alpha_3^3M_B^4}\rho^2_3(\alpha_3)\} e^{-s/M_B^2} \nonumber\\& -\frac{1}{2}\frac{1}{\alpha_{30}^2 M_{\Xi}^2-q^2+m_c^2}[\rho^2_2(\alpha_{30})-\frac{2}{\alpha_{30} M_B^2}\rho^2_3(\alpha_{30})]e^{-s_0/M_B^2} \nonumber\\& -\frac{\alpha_{30}^2}{\alpha_{30}^2 M_{\Xi}^2-q^2+m_c^2}[\frac{d}{d\alpha_{30}}\frac{\rho^2_3(\alpha_{30})}{\alpha_{30}(\alpha_{30}^2 M_{\Xi}^2-q^2+m_c^2)}] e^{-s_0/M_B^2},
   \end{align}
  
  where the $\rho^2_i(i=1,2,3)$ are
  
   \begin{align}
    \rho^2_1(\alpha_3)=&B_1(\alpha_3)M_\Xi,\\  
    \rho^2_2(\alpha_3)=&[2B_2(\alpha_3)+B_4(\alpha_3)-B_5(\alpha_3)]M_\Xi^2m_c,\\  
    \rho^2_3(\alpha_3)=&B_6(\alpha_3)M_\Xi^3m_c^2.
  \end{align}
  
     Form factor $g_1(q^2)$
   \begin{align}
    f_{\Omega_c}g_1 e^{-M_{\Omega_c}^2/M_B^2}&=\int_{\alpha_{30}}^{1}d\alpha_3 \{\frac{1}{\alpha_3}\rho^3_1(\alpha_3)+\frac{1}{\alpha_3^2 M_B^2}\rho^3_2(\alpha_3,q^2)-\frac{1}{\alpha_3^3 M_B^4}\rho^3_3(\alpha_3)\}e^{-s/M_B^2} \nonumber\\ & +\frac{1}{\alpha_{30}^2 M_{\Xi}^2-q^2+m_c^2}[\rho^3_2(\alpha_{30},q^2)-\frac{1}{\alpha_{30}M_B^2}\rho^3_3(\alpha_{30})]e^{-s_0/M_B^2} \nonumber \\ & +\frac{\alpha_{30}^2}{\alpha_{30}^2 M_{\Xi}^2-q^2+m_c^2}[\frac{d}{d\alpha_{30}}\frac{\rho^3_3(\alpha_{30})}{\alpha_{30}(\alpha_{30}^2 M_{\Xi}^2-q^2+m_c^2)}]e^{-s_0/M_B^2},
   \end{align}

    where $\rho^3_i(i=1,2,3)$ are
    \begin{align}
    \rho^3_1(\alpha_3)=&B_1(\alpha_3)m_c-\alpha_3 B_3(\alpha_3)M_{\Xi},\\
    \rho^3_2(\alpha_3,q^2)=&B_2(\alpha_3)[\alpha_3(\alpha_3-1)M_\Xi^3-\alpha_3M_\Xi q^2+\alpha_3M_\Xi(M_{\Omega_c}^2+M_\Xi m_c)-M_\Xi m_c^2]\nonumber \\&-\frac{1}{2}\alpha_3B_4(\alpha_3)M_\Xi^2m_c+\frac{1}{2}B_5(\alpha_3)(2\alpha_3^2M_\Xi^3-\alpha_3M_\Xi^2m_c),\\
    \rho^3_3(\alpha_3)=&\alpha_3^2B_6(\alpha_3)M_\Xi^4m_c.
    \end{align}
  
       Form factor $g_2(q^2)$
     \begin{align}
   f_{\Omega_c}(g_1 &  -\frac{M_{\Omega_c}-M_\Xi}{M_{\Omega_c}}g_2)e^{-M_{\Omega_c}^2/M_B^2}=\int_{\alpha_{30}}^{1}d\alpha_3 \{\frac{1}{\alpha_3}\rho^4_1(\alpha_3) \nonumber\\& -\frac{1}{\alpha_3^2 M_B^2}\rho^4_2(\alpha_3)+\frac{1}{\alpha_3^3 M_B^4}\rho^4_3(\alpha_3)\}e^{-s/M_B^2} \nonumber\\& -\frac{1}{\alpha_{30}^2 M_{\Xi}^2-q^2+m_c^2}[\rho^4_2(\alpha_{30}) -\frac{1}{\alpha_{30}M_B^2}\rho^4_3(\alpha_{30})] \nonumber\\& -\frac{\alpha_{30}^2}{\alpha_{30}^2 M_\Xi^2-q^2+m_c^2}[\frac{d}{d\alpha_{30}}\frac{\rho^4_3(\alpha_{30})}{\alpha_{30}(\alpha_{30}^2 M_\Xi^2-q^2+m_c^2)}]e^{-s_0/M_B^2}.
    \end{align}  
    	
    where $\rho^4_i(i=1,2,3)$ are
    \begin{align}
    	\rho^4_1(\alpha_3)=&B_1(\alpha_3)M_\Xi,\\
    	\rho^4_2(\alpha_3)=&[B_2(\alpha_3)+\frac{1}{2}B_4(\alpha_3)-\frac{1}{2}B_5(\alpha_3)]M_\Xi^2m_c,\\
    	\rho^4_3(\alpha_3)=&B_6(\alpha_3)M_\Xi^3m_c^2.
    	\end{align}

    	The $\alpha_{30}$ relates to the threshold $s_0$ is	

	\begin{gather}
	\alpha_{30}=\frac{-(s_0-M_\Xi^2-q^2)+\sqrt{(s_0-M_\Xi^2-q^2)^2-4M_\Xi^2(q^2-m_c^2)}}{2M_\Xi^2}
	\end{gather}
%	\end{widetext}
	
	The Borel transform parameters is
	\begin{gather}
	s=(1-\alpha_3)M_\Xi^2-\frac{1-\alpha_3}{\alpha_3}q^2+\frac{m_c^2}{\alpha_3}
	\end{gather}
	
	The signs $B_i(\alpha_3)(i=1\cdots6)$ we used above are defined in the following expressions
	\begin{align}
			B_1(\alpha_3)=&\int_{0}^{1-\alpha_3}d\alpha_1 V_1(\alpha),\\
				B_2(\alpha_3)=&\int_{0}^{1-\alpha_3}d\alpha_3'\int_{0}^{1-\alpha_3'}d\alpha_1(V_1-V_2-V_3)(\alpha'),\\
				B_3(\alpha_3)=&\int_{0}^{1-\alpha_3}d\alpha_1V_3(\alpha),\\
				B_4(\alpha_3)=&\int_{0}^{\alpha_3}d\alpha_3'\int_{0}^{1-\alpha_3'}d\alpha_1(-2V_1+V_3+V_4+2V_5)(\alpha'),\\
				B_5(\alpha_3)=&\int_{0}^{\alpha_3}d\alpha_3'\int_{0}^{1-\alpha_3'}d\alpha_1(V_4-V_3)(\alpha'),\\
				B_5(\alpha_3)=&\int_{0}^{\alpha_3}d\alpha_3'\int_{0}^{\alpha_3'}d\alpha_3''\int_{0}^{1-\alpha_3''}d\alpha_1(-V_1+V_2 +V_3+V_4+V_5-V_6)(\alpha''),
    \end{align}
    where $(\alpha)=(\alpha_1,1-\alpha_1-\alpha_3,\alpha_3), (\alpha_3')=(\alpha_1,1-\alpha_1-\alpha_3',\alpha_3'), (\alpha_3'')=(\alpha_1,1-\alpha_1-\alpha_3'',\alpha_3'')$.

	\section{Numerical Analysis.}  \label{sec:III}
    
       \begin{figure*}[t]
		\includegraphics[width = 0.4\textwidth]{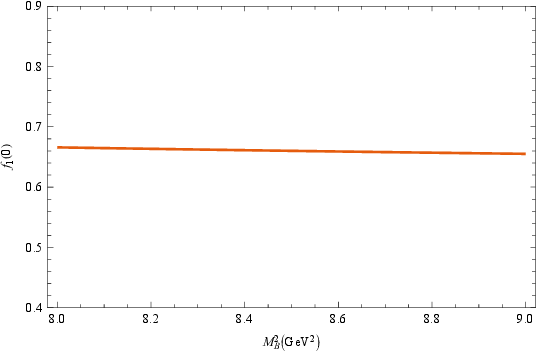}\qquad
		\includegraphics[width=0.4\textwidth]{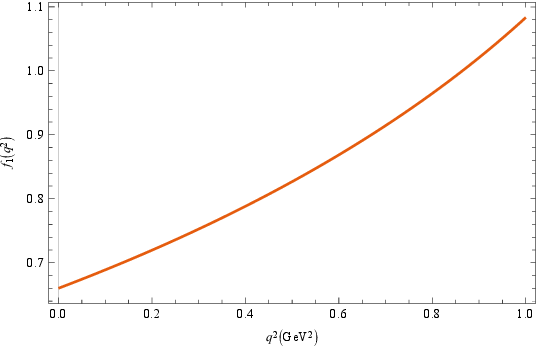}	
    	\caption{The dependence of form factor $f_1$ on Borel Parameter at $q^2=0$ between $8{\rm{GeV}}^2 < M_B^2 <9{\rm{GeV}}^2$ (left), and the form factor $f_1$ at $M_B^2=8.5 {\rm{GeV}}^2$ of the momentum transfer square $0<q^2<1 {\rm{GeV}}^2$ (right). } \label{fig1}
    \end{figure*}
    
	In order to know the form factors and give the value properties of the semileptonic decay of $\Omega_c \to \Xi^- l^+ \nu_l$, the numerical value of the parameters in the formulas of form factors should be set. In this analysis, we adopt the standard center values of charm quark mass and baryons masses from PDG~\cite{Zyla:2020}, which gives the numbers $m_c=1.27~{\rm{GeV}}, M_{\Xi^-}=1.3217~{\rm{GeV}}$, and the $\Omega_c$ baryon mass $M_{\Omega_c^0}=2.6952~{\rm{GeV}}$. The physical region of momentum transfer square $q^2$ varies in the region $0<q^2<(M_{\Omega_c^0}-M_\Xi)^2$. For the numerical analysis, we also need to know the nonperturbative parameters of the $\Omega_c$ baryon decay constants and the $\Xi$ baryon decay constants $f_\Xi$ and $\lambda_1$. The nonpertubative parameters distribution amplitudes $V_i (i=1,...,8)$ have been derived in ref.~\cite{Liu:2009lc} where we list them in the appendix.

    In this calculation, the decay constant of $\Omega_c$ baryon which adopted has been calculated with full QCD sum rule theory from ref.~\cite{Wang:2010}, the value of which is $f_{\Omega_c}=(0.093\pm0.023)~{\rm{GeV}}^3$, which is in accordance with the result calculated from ref.~\cite{Zhang:2008} where should be $0.08~{\rm{GeV}}^3$. This is different from that given by reference~\cite{Chen:2017,Liu:2008}, where they calculated the decay constant of $\Omega_c$ by the method QCD sum rules in the framework of heavy quark effective theory. The decay constant of $\Xi$ can be found in~\cite{Liu:2009lc}, the values $f_\Xi=(9.9\pm0.4)\times10^{-3}~{\rm{GeV}}^2, \lambda_1=-(2.8\pm0.1)\times10^{-2}~{\rm{GeV}}^2$.

\begin{figure*}[!]
			\begin{center}
				\includegraphics[width=0.4\textwidth]{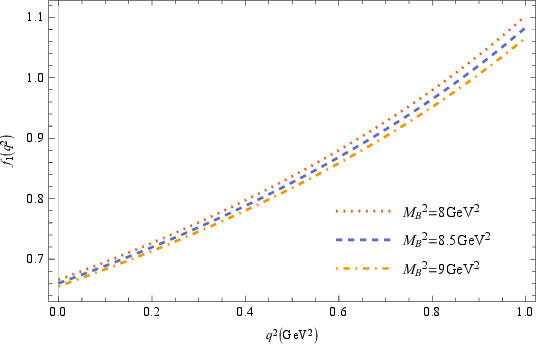}
				\qquad
				\includegraphics[width=0.4\textwidth]{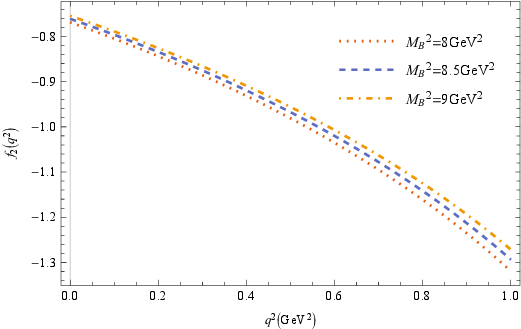}\\
				\includegraphics[width=0.4\textwidth]{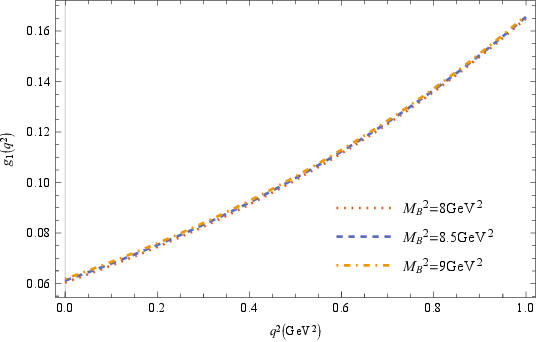}
				\qquad
				\includegraphics[width=0.4\textwidth]{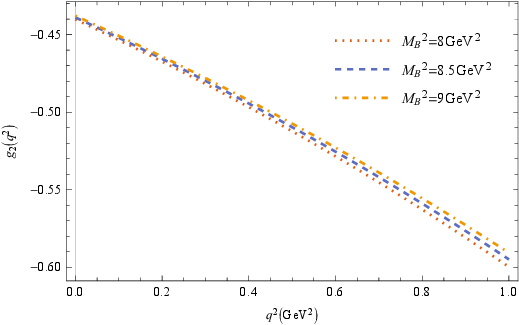}
			\end{center}
			\caption{The form factors $f_i(i=1,2)$ and $g_i(i=1,2)$ vary from zero momentum transfer square to $q^2=1~{\rm{GeV}}^2$} \label{fig2}
		\end{figure*}

	The other parameters which should be determined are the Borel region. The chosen principle of Borel parameters is that we should promise to suppress both the higher resonance and twist contributions. In ~\cite{Zhang:2008,Wang:2010}, the QCD sum rule method has been used to estimate the mass of $\Omega_c$ baryon. In ref~\cite{Zhang:2008,Wang:2010}, the chosen threshold of the $\Omega_c$ charm baryon around the first excited states was $\sqrt{s_0}=3.3~{\rm{GeV}}\sim3.5~{\rm{GeV}}$, and the mass was calculated at $\sqrt{s_0}=3.4~{\rm{GeV}}$. So the threshold of $s_0$ we choose is $\sqrt{s_0}=3.4~{\rm{GeV}}$ in our calculations for accordance, and the stable window of the  Borel parameter $M_B^2$ working we analyse is $8<M_B^2<9~{\rm{GeV}}^2$. We analyse the varying of $M_B^2$ at the point $\sqrt{s_0}=3.4$ GeV and $q^2=0~{\rm{GeV}}^2$. The pictures of curve are given in Fig.~\ref{fig1}, and the pictures of form factors $f_i (g_i)(i=1,2)$ are plotted in Fig.~\ref{fig2}.

	The decay width of the semileptonic process should be investigated on the whole physical region, while the light-cone QCD sum rules only worked in the region $q^2\ll (M_{\Omega_c}-M_\Xi)^2$, so the form factors are studied varying from $0~{\rm{GeV}}^2$ to $1~{\rm{GeV}}^2$. So that, we extrapolate these form factors by the three-parameter dipole formula Eq.~(\ref{dipole formula}) on the whole kinematical region $0<q^2<(M_{\Omega_c}-M_\Xi)^2$. Another benefit of using this fitting formula is that the procedure of calculation can be simplified. 
	 
	 \begin{gather}
	 f_i(q^2)=\frac{f_i(0)}{a(q^2/M_{\Omega_c}^2)^2+b(q^2/M_{\Omega_c}^2)+1}. \label{dipole formula}
	 \end{gather}
	 	
	 	\begin{table}[!]
	 		\centering
	 		\caption{The form factors of transition from $\Omega_c$ to $\Xi$ at momentum transfer $q^2=0~{\rm{GeV}}^2$.The comparison between our work and others' work are also showed in this table.} \label{table 1}
	 		\begin{tabular}{{|c|c|c|c|c|c|}}\hline
	 			$f_i(0)$&This work&NRQM~\cite{Rerez:1989}&HQET~\cite{Chen:1997}&LFQM~\cite{Zhao:2018}&MIT bag~\cite{Hu:2020}\\\hline
	 			$f_1(0)$&$0.66\pm0.02$&-0.23&-0.34&0.653&0.34\\\hline
	 			$f_2(0)$&$-0.76\pm0.03$&0.21&0.35&0.620&-\\\hline
	 			$g_1(0)$&$0.06\pm0.01$&0.14&0.10&-0.182&-0.15\\\hline
	 			$g_2(0)$&$-0.44\pm0.01$&-0.019&-0.020&0.002&-\\\hline
	 		\end{tabular}
	 	\end{table}
	 
	  \begin{table}[h]
	  	\centering
	  	\caption{The form factors at threshold value $\sqrt{s_0}=3.4~{\rm{GeV}}$ and Borel parameter $M_B^2=8.5~{\rm{GeV}}^2$ and relative coefficients fitted from dipole formula in Eq.~(\ref{dipole formula}).} \label{table 2}
	  	\begin{tabular}{{|c|c|c|c|}}\hline
	  		$f_i$&$f_i(0)$&{\it a}&{\it b}\\\hline
	  		$f_1$&$0.66$&1.28&-3.01\\\hline
	  		$f_2$&$-0.76$&1.58&-3.2\\\hline
	  		$g_1$&$0.06$&16.99&-6.9\\\hline
	  		$g_2$&$-0.44$&1.62&-2.13\\\hline
	  	\end{tabular}
	  \end{table}
	  	
	 The form factors at point $q^2=0~{\rm{GeV}}^2$ are given in Table~\ref{table 1}, and the results obtained from other approaches are also listed in this table. The coefficients {\it a} and {\it b} we fitted are listed in Table~\ref{table 2}. We use the decay width of semileptonic decay as that in~\cite{Huang:2004,Liu:2009lc}. The differential decay rate dependent on $q^2$ is 
	 
	 \begin{align}
	 \frac{d\Gamma}{dq^2}=&\frac{G_F^2|V_{cd}|^2}{192\pi^3M_{\Omega_c}^5}q^2\sqrt{q_+^2q_-^2}\{-6f_1f_2M_{\Omega_c}m_+q_-^2+6g_1g_2M_{\Omega_c}m_-q_+^2+f_1^2M_{\Omega_c}^2(\frac{m_+^2m_-^2}{q^2}+m_-^2-\nonumber\\&2(q^2+2M_{\Omega_c}M_\Xi))+g_1^2M_{\Omega_c}^2(\frac{m_+^2m_-^2}{q^2}+m_+^2-2(q^2-2m_{\Omega_c}M_\Xi))-f_2^2[-2m_+^2m_-^2+m_+^2q^2\nonumber\\&+q^2(q^2+4M_{\Omega_c}M_\Xi)]-g_2^2[-2m_+^2m_-^2+m_-^2q^2+q^2(q^2-4M_{\Omega_c}M_\Xi)]\}.
	 \end{align}
	 
\begin{figure}[h]
    	\begin{center}
    		\includegraphics[width = 0.4\textwidth]{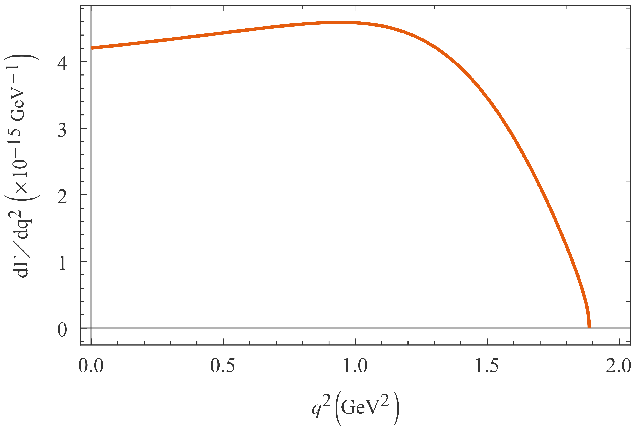}	
    	\end{center}
    	\caption{The differential decay rate dependence on $q^2$.} \label{fig3}
    \end{figure}

	 In the expressions $m_\pm=M_{\Omega_c}\pm M_\Xi$ and $q_{\pm}^2=q^2-m_\pm^2$. The fermi constant is $G_F=1.166\times10^{-5}~{\rm{GeV}}^{-2}$, and CKM matrix element is $|V_{cd}|=0.221$. Substituting these constants into the differential decay formula and integrating it on the whole dynamical region, the decay width of the weak semileptonic decay $\Omega_c^0 \to \Xi^- l^+ \nu_l$ can be obtained, that is $\Gamma=(7.51\pm0.36)\times 10^{-15}~{\rm{GeV}}$. The differential decay width is plotted in Fig~\ref{fig3}. With the mean lifetime of $\Omega_c^0$ in PDG~\cite{Aaij:2018,Zyla:2020}, which gives the value $\tau=268\times10^{-15}s$, the branching ratio of decay which we estimate by the dipole formula is $Br(\Omega_c^0 \to \Xi^- l^+ \nu_l)=(3.06\pm0.15)\times10^{-3}$. The errors of form factors, decay width and branching ratio come from the threshold $\sqrt{s_0}$ and Borel parameter $M_B$. We also list the results calculated by other methods compared with ours in Table~\ref{table 3}.	 	

         \begin{table}[h]
        	\centering
        	\caption{The decay widths and branching ratio which we calculate compared with other methods.} \label{table 3}
        	\begin{tabular}{{|c|c|c|c|}}\hline
 	     	Decay channel&$\Gamma / {\rm{GeV}}$&$\mathcal{B}$&Reference\\\hline
 	    	$\Omega_c^0\to\Xi^-l^+\bar{\nu}_l$&$(7.51\pm0.36)\times10^{-15}$&$(3.06\pm0.15)\times10^{-3}$&our work\\\hline
 	    	$\Omega_c^0\to\Xi^-e^+\nu_e$&$2.08\times10^{-15}$&$2.18\times10^{-4}$&\cite{Zhao:2018}\\\hline
        	\end{tabular}
        \end{table}
        
        The form factors calculated in this work are comparable with other calculations in different methods, the decay width in light-cone QCD sum rule calculation is in accordance with the light-front approach in the same order. But there is no experiment result has been reported, so only the test in future experiments can confirm these values. 
    
	\section{Conclusions}  \label{sec:IV}
	The process of the semileptonic decay from heavy baryon $\Omega_c$ to light baryon $\Xi$ and two leptons is calculated with light-cone sum rule approach. With the light-cone sum rules, the form factors of this weak decay are calculated and the explicit expressions of them are given. By using the numerical values in PDG the form factors are calculated and their numerical values are given in Table~\ref{table 1}. The form factors obtained by other approaches are compatible with ours in Table~\ref{table 1} too. The decay width of $\Omega_c^0 \to \Xi^-l^+\nu_l$ is evaluated to the number $\Gamma=(7.51\pm0.36)\times10^{-15}~{\rm{GeV}}$. It's larger than the result given in ref.~\cite{Zhao:2018}, the reason may be the different values of form factors come from different approaches. The branching ratio of it is given by  $Br(\Omega_c^0 \to \Xi^- l^+ \nu_l)=(3.06\pm0.15)\times10^{-3}$. But there is no absolute branching of $\Omega_c^0$ decay which has been discovered in experiments, and the only reference of it is to set the $\Omega_c^0 \to \Omega^- \pi^+$ decay channel to one, and the other decay channels are all relative to it~\cite{Frabetti:1993}. Due to the branching ratio is in the order ten to minus three, this calculation results will give a prediction of this decay channel and it may be an easy observed decay channel in experiment, and will also be tested in the future experiments.
	\section*{Acknowledgments}
	This work was supported in part by the National Natural Science Foundation of China under Contract No. 11675263.
	\section*{Appendix}  \label{appendix}
	\appendix
	\setcounter{equation}{0}
	\renewcommand\theequation{A.\arabic{equation}}
	In this appendix we give the distribution amplitudes of $\Xi$ baryon~\cite{Liu:2009lc}.
	\begin{align}
		\langle 0|\epsilon_{ijk} s_\alpha^{iT}(0) s_\beta^j(0)d_\gamma^k(x)|\Xi(p)\rangle=& \mathcal{V}_1(\slashed{p}C)_{\alpha \beta}(\gamma_5 \Xi)_\gamma+\mathcal{V}
	_2 M_\Xi (\slashed{p}C)_{\alpha \beta}(\slashed{x}\gamma_5 \Xi)_\gamma+\mathcal{V}_3 M_\Xi (\gamma_\mu C)_{\alpha\beta}(\gamma^\mu \gamma_5\Xi)_\gamma+\nonumber\\&\mathcal{V}_4 M_\Xi^2(\slashed{x}C)_{\alpha\beta}(\gamma_5\Xi)_\gamma+\mathcal{V}_5 M_\Xi^2(\gamma_\mu C)_\alpha\beta(i\sigma^{\mu\nu}x_\nu\gamma_5\Xi)_\gamma+\mathcal{V}_6 M^3(\slashed{x}C)_{\alpha\beta}(\slashed{x}\gamma_5\Xi)_\gamma
    \end{align}
	
	The distribution amplitudes $\mathcal{V}_i, (i=1,...,6)$ did not have definite twist, in order to get the distribution amplitudes with the definite twist we should make the following transforming to get the amplitudes with definite twist
	
	\begin{align}		
	\mathcal{V}_1&=V_1, \quad 2p\cdot x \mathcal{V}_2=V_1-V_2-V_3, \nonumber\\
	2\mathcal{V}_3&=V_3, \quad   4p\cdot x\mathcal{V}_4=-2V_1+V_3+V_4+2V_5,\nonumber\\
	4p\cdot x\mathcal{V}_5&=V_4-V_3,\nonumber\\ 
	(2p\cdot x)^2\mathcal{V}_6&=-V_1+V_2+V_3+V_4+V_5-V_6.	
	\end{align}
	
	These distribution amplitudes $V_i$ are functions of $x\cdot p$, but we need the variable relevant to the longitude momentum fraction of quarks in the baryon, so we should make the following transformation formula.
	\begin{equation}
	F(\alpha_i p \cdot z)=\int \mathcal{D}xe^{-ipx \sum_{i}x_i\alpha_i}F(x_i),
	\end{equation}
	The integration measure $\int \mathcal{D}x$ is
	\begin{equation}
	\int \mathcal{D}x=\int_{0}^{1}dx_1dx_2dx_3\delta(x_1+x_2+x_3-1).
	\end{equation}
	The distribution amplitudes of $\Xi$ with twist-3 is
	\begin{equation}
	V_1(x_i)=120x_1x_2x_3\phi_3^0. \label{twist-3 DA}
	\end{equation}
	Twist-4 distribution amplitudes
  \begin{equation}
  V_2(x_i)=24x_1x_2\phi_4^0, \quad  V_3(x_i)=12x_3(1-x_3)\psi_4^0.
  \end{equation}
  Twist-5 distribution amplitudes
  \begin{equation}
  V_4(x_i)=3(1-x_3)\psi_5^0, \qquad   V_5(x_i)=6x_3\phi_5^0.
  \end{equation}
  And twist-6 distribution amplitudes
  \begin{equation}
  V_6(x_i)=2\phi_6^0.   \label{twist-6 DA}
  \end{equation}
  In the leading order conformal spin accuracy, the coefficients of (\ref{twist-3 DA})-(\ref{twist-6 DA}) can be expressed as
  \begin{align}
  &\phi_3^0=\phi_6^0=f_\Xi, \\ &\phi_4^0=\phi_5^0=\frac{1}{2}(f_\Xi+\lambda_1), \\ &\psi_4^0=\psi_5^0=\frac{1}{2}(f_\Xi-\lambda_1).
  \end{align}

\end{document}